\documentclass{raa}

\usepackage{graphicx,times}
\usepackage{natbib}
\usepackage{amssymb,amsmath}
\usepackage{placeins}
\usepackage{url}
\bibpunct{(}{)}{;}{a}{}{,}
\usepackage[pagebackref=true]{hyperref}

\begin{document}

\title{Orbital and spin periods of the intermediate polar 2PBC J0800.5$-$4306}

\volnopage{Vol.0 (20xx) No.0, 000--000}
\setcounter{page}{1}

\author{Yabing Zhao\inst{1}
\and Chunhua Zhu\inst{1}
\and Sufen Guo\inst{1,2,3,4}
\and Guoliang L\"{u}\inst{1}
\and Xizhen Lu\inst{1}
\and Xinyu Liu\inst{1}}

\institute{School of Physical Science and Technology, Xinjiang University, Urumqi 830046, China; {\it chunhuazhu@sina.cn; guosufen@xju.edu.cn; guolianglv@sina.com}\\
\and Yunnan Observatories, Chinese Academy of Sciences (CAS), Kunming 650216, People's Republic of China\\
\and Key Laboratory for the Structure and Evolution of Celestial Objects, CAS, Kunming 650216, People's Republic of China\\
\and International Centre of Supernovae, Yunnan Key Laboratory, Kunming 650216, People's Republic of China\\
}

\abstract{
2PBC J0800.5$-$4306 (Swift J0800.7$-$4309; optical counterpart Gaia DR3 5533020382479508736) is a hard-X-ray-selected cataclysmic variable whose magnetic subtype had remained uncertain. We analyse \textit{TESS} photometry from Sectors 88 and 89 and archival \textit{XMM-Newton} timing data. The \textit{TESS} data show signals at the orbital frequency $\Omega$, its first harmonic $2\Omega$, the white-dwarf spin frequency $\omega$, and the spin--orbit beat frequency $\omega-\Omega$. We measure $P_{\rm orb}=5.38540^{+0.00027}_{-0.00028}$ h and $P_{\rm spin}=0.7015570^{+0.0000059}_{-0.0000062}$ h. The \textit{XMM-Newton} EPIC-pn, MOS1, and MOS2 periodograms show broad peaks near the 2525.6-s (0.701557-h) optical spin period. Together with the previously published cataclysmic variable classification and Swift/BAT hard-X-ray detection, these timing results identify 2PBC J0800.5$-$4306 as an intermediate polar. We report the first published measurements of its orbital and white-dwarf spin periods, while SPOC difference-image diagnostics and pixel-level timing tests support that the measured \textit{TESS} modulations are associated with the Gaia counterpart despite the crowded field.
\keywords{novae, cataclysmic variables --- stars: individual: 2PBC J0800.5$-$4306 --- stars: magnetic field --- stars: rotation --- white dwarfs --- X-rays: binaries}
}

\authorrunning{Y. Zhao, C. Zhu, S. Guo et al.}
\titlerunning{Periods of 2PBC J0800.5$-$4306}
\maketitle

\section{Introduction}

Accreting white-dwarf (WD) binaries serve as important astrophysical laboratories for studying mass transfer, accretion physics, and compact-binary evolution. Examples include white-dwarf--Be-star binaries, symbiotic stars, and cataclysmic variables, which represent different combinations of donor stars, mass-transfer modes, and evolutionary pathways \citep{1995cvs..book.....W,2023RAA....23b5021Z,2025ApJ...995...14Z}. Among them, cataclysmic variables (CVs) are semi-detached binaries in which a WD accretes matter from a Roche-lobe-filling donor, usually a low-mass, late-type star \citep{1995cvs..book.....W,2017PASP..129f2001M}. They display a wide range of photometric and spectroscopic variability, including flickering, dwarf-nova outbursts, nova eruptions, eclipses, and coherent periodic modulations.

The observational census of CVs has expanded substantially through large-scale spectroscopic, astrometric, time-domain, and X-ray surveys. \citet{2020AJ....159...43H} spectroscopically identified 245 CVs in LAMOST data, including 58 new candidates, while \citet{2023MNRAS.524.4867I} compiled 507 CVs observed by SDSS, including 70 new classifications and 59 new periods. The latter study also identified 18 objects that had previously been classified incorrectly as CVs, illustrating the importance of multiwavelength verification. Accurate \textit{Gaia} parallaxes have enabled the construction of volume-limited samples and more reliable measurements of CV space densities and population properties \citep{2020MNRAS.494.3799P}. More recently, \citet{2024A&A...686A.110S} combined eROSITA X-ray detections with SDSS spectroscopy to identify and classify accreting WD binaries.

The secular evolution of CVs involves angular-momentum loss, mass transfer, and the thermal response of the donor star. In standard evolutionary models, gravitational radiation is usually adopted as the dominant angular-momentum loss mechanism below the orbital-period gap, whereas magnetic braking is adopted above it. Comparisons between observed donor star mass--radius relations and binary evolution models can constrain the effective long-term mass transfer and angular-momentum loss rates \citep{2011ApJS..194...28K}. Nova eruptions also contribute to Galactic chemical evolution. Nova evolution calculations combined with population synthesis indicate that novae can be an important source of Galactic lithium \citep{2024ApJ...971....4G}. Similar nova evolution and population synthesis calculations show that nova ejecta can contribute substantially to the Galactic abundances of the odd-mass isotopes $^{13}$C, $^{15}$N, and $^{17}$O \citep{2024RAA....24j5007H}. CVs therefore provide important tests of accretion-disc physics, thermonuclear runaways, angular-momentum loss, and WD mass and spin evolution.

A particularly important subset of CVs is the magnetic cataclysmic variables (mCVs), in which the WD magnetic field governs the structure of the accretion flow. In the volume-limited sample of 42 CVs within 150 pc assembled by \citet{2020MNRAS.494.3799P}, mCVs constitute approximately 36 per cent of the observed systems. The two principal subclasses of mCVs are polars and intermediate polars (IPs). In polars, the WD rotation is approximately synchronized with the binary orbit, and the magnetic field prevents the formation of a full accretion disc. In IPs, the WD rotates asynchronously, usually substantially faster than the binary orbit. The magnetic field truncates the inner accretion disc and channels the gas along field lines towards the magnetic poles \citep{1994PASP..106..209P,2017PASP..129f2001M}.

Modern multiwavelength observations have improved the identification and physical characterization of both subclasses. For polars, cyclotron harmonics, circular polarization, and time-resolved spectroscopy can directly reveal magnetic accretion and constrain the WD magnetic field. For example, \citet{2025A&A...694A.112L} identified ZTF J0112+5827 as a polar with an orbital period of 80.9 min and estimated a WD magnetic-field strength of approximately 38.7 MG from its cyclotron harmonics. Their Doppler tomography revealed accretion streams but no evidence for an accretion disc. In IPs, magnetically channelled gas forms a standing shock above the WD surface. As the post-shock plasma settles towards the surface, it cools from temperatures of tens of keV and produces a multi-temperature, optically thin X-ray spectrum. This hot plasma accounts for the prominent hard-X-ray emission of many IPs \citep{2017PASP..129f2001M,2022MNRAS.511.4937S}.

Coherent modulation at the white-dwarf spin frequency $\omega$ is a central observational signature of an intermediate polar (IP) and can be detected in both optical and X-ray data. Optical power spectra may also contain a distinct spin--orbit beat signal at $\omega-\Omega$, whose frequency depends on both the white-dwarf spin and the binary orbital motion \citep{2017PASP..129f2001M}. \citet{Bruch_2025} analysed light curves obtained with the Transiting Exoplanet Survey Satellite (\textit{TESS}; \citealt{2015JATIS...1a4003R}) for 121 IPs and candidates and detected periodic photometric signals attributed to the white-dwarf spin in approximately half of the systems.

Theoretically, magnetic-accretion models predict rotational equilibria across the $P_{\rm spin}/P_{\rm orb}$--$\mu_1$ plane, where $\mu_1$ is the white-dwarf magnetic moment. Different regions of this parameter space are associated with disc-like, stream-like, and ring-fed accretion flows \citep{Norton_2004}. Recent population-synthesis calculations show that allowing strong white-dwarf magnetic fields to appear once the white dwarf reaches a specified age can reproduce the observed fraction of magnetic CVs and its dependence on orbital period \citep{2025A&A...698L..22S}, while calculations using different magnetic-braking prescriptions show that the adopted braking law strongly affects the predicted mass-transfer rates, orbital-period distribution, and formation of the period gap \citep{2026A&A...707A..76Z}. Nevertheless, the origin and evolution of WD magnetic fields, the detailed accretion geometry of many IPs, and the evolutionary relationship between IPs and polars remain uncertain. Many candidate IPs also lack securely measured orbital and spin periods. Detailed optical and X-ray timing studies are therefore essential for confirming individual systems and providing the parameters required to test magnetic-accretion and spin-evolution models.

2PBC J0800.5$-$4306 appeared in the second Palermo \textit{Swift}/BAT catalogue without an identified counterpart or source classification \citep{2010A&A...524A..64C}. \citet{Rojas2017} subsequently identified its optical counterpart as a CV through optical spectroscopy. From the equivalent-width ratio of He\,II $\lambda4686$ to H$\beta$, they suggested that the system might be a polar, while emphasizing that time-resolved follow-up was required to confirm its magnetic nature. Using \textit{Swift}/XRT imaging, \citet{Landi2017} detected three X-ray sources within the 90 per cent IBIS/BAT positional uncertainties. They favoured the brightest source, which was also the brightest above 3 keV and coincided with the optically classified CV, as the counterpart of Swift J0800.7$-$4309. Its measured 2--10 keV flux was $(5.0\pm0.2)\times10^{-12}$ erg cm$^{-2}$ s$^{-1}$; however, the authors noted that a contribution from the second XRT source could not be completely excluded without optical spectroscopy of that source. Mukai's Intermediate Polar Home Page\footnote{\url{https://asd.gsfc.nasa.gov/Koji.Mukai/iphome/systems/pbc0800.html}} lists the object as a possible IP. The optical counterpart is Gaia DR3 5533020382479508736 (hereafter Gaia 5533), which also appears in hot-subluminous-star catalogues \citep{2022A&A...662A..40C,2025A&A...693A.268R} because of its location in the \textit{Gaia} colour--magnitude diagram.

In this work, we analyse \textit{TESS} photometry and archival \textit{XMM-Newton} observations to determine the orbital and white-dwarf spin periods of 2PBC J0800.5$-$4306 and to establish its IP nature. Because the target lies in a crowded field, we also use SPOC difference-image diagnostics and pixel-level timing analysis to test whether the periodic \textit{TESS} signals originate from its Gaia counterpart, Gaia 5533. Section~\ref{sec:data} describes the observations, Section~\ref{sec:methods} presents the timing and localization methods, Section~\ref{sec:results} gives the measured periods and IP classification, and Section~\ref{sec:conclusion} summarizes our conclusions.

\section{Data}
\label{sec:data}

\subsection{TESS photometry and target-pixel files}

Gaia 5533 was observed by \textit{TESS} in Sectors 88 and 89. We retrieved the 20-s and 2-min Science Processing Operations Center (SPOC) light-curve and target-pixel files from the Mikulski Archive for Space Telescopes (MAST) with \texttt{lightkurve} \citep{2018ascl.soft12013L}. The primary timing analysis used the 2-min \texttt{PDCSAP\_FLUX} light curves from Sectors 88 and 89. For each sector, we retained cadences with \texttt{QUALITY}=0 and valid time and flux measurements, divided the \texttt{PDCSAP\_FLUX} values by the median flux of that sector, and removed slow variations with a 1-d biweight filter implemented in \texttt{wotan} \citep{2019AJ....158..143H}. The two independently normalized and detrended sector light curves were then concatenated for the primary period analysis. The 20-s \texttt{PDCSAP\_FLUX} and \texttt{SAP\_FLUX} products were analysed separately as cadence- and flux-product consistency checks; they were not combined with the 2-min measurements.

The target is faint ($T\simeq16.90$) and lies in a crowded field. The SPOC headers give \texttt{CROWDSAP}=0.0986 for Sector 88 and 0.0521 for Sector 89. \texttt{CROWDSAP} estimates the fraction of the mean flux in the adopted aperture that is contributed by the target. These values imply substantial dilution and prevent us from interpreting the normalized aperture amplitude as an undiluted intrinsic amplitude. They do not, by themselves, identify which star produces a periodic change; that question is addressed with difference images and pixel-level timing in Section~\ref{sec:localization}.

\subsection{XMM-Newton data}

We used archival \textit{XMM-Newton} observation 0820330901 \citep{2001A&A...365L...1J}. We analysed the PPS background-subtracted \texttt{RATE} products for the X-ray source positionally coincident with Gaia 5533 (PPS source identifier 1) from the EPIC-pn, MOS1, and MOS2 cameras in the broad 0.2--12 keV band. Its astrometrically corrected XMM catalogue position, ${\rm RA}=120.166593^\circ$ and ${\rm Dec}=-43.185475^\circ$, is separated from the Gaia 5533 position by only 0.254 arcsec. The three source light curves therefore correspond to the X-ray source positionally associated with the optical counterpart. The \texttt{RATE} products contain the standard pipeline source and background good-time intervals (GTIs). We did not reprocess the event files or apply an additional manual high-background cut, so the analysis inherits the PPS GTI filtering. The source-extraction radius was 40 arcsec in all three products; no other catalogue EPIC source position lies within this radius. We retained rows with valid \texttt{TIME}, background-subtracted \texttt{RATE}, and \texttt{ERROR} values, requiring \texttt{ERROR}$>0$; negative \texttt{RATE} values were retained because a background-subtracted count rate can be negative as a statistical measurement. The retained light curves contain 10,804, 870, and 850 time bins for EPIC-pn, MOS1, and MOS2, respectively. The \texttt{TIMEDEL} header values give time resolutions of 1.46, 30, and 30 s, while the first-to-last retained-bin spans are 24.14, 26.10, and 25.47 ks. For these retained bins, $\sum(\texttt{FRACEXP}\times\texttt{TIMEDEL})$ gives exposure times of 13.63, 25.67, and 25.10 ks, respectively.

\section{Methods}
\label{sec:methods}

\subsection{Period analysis}

We computed generalized Lomb--Scargle periodograms of the detrended \textit{TESS} flux, allowing a constant offset (a floating mean) to be fitted at each trial frequency and using the standard power normalization \citep{1976Ap&SS..39..447L,1982ApJ...263..835S,2009A&A...496..577Z}. The displayed periodogram uses the full concatenated 2-min \texttt{PDCSAP\_FLUX} time series. We searched from 1 to 45 d$^{-1}$ for the orbital, spin, harmonic, and beat-frequency structure. We denote the orbital frequency by $\Omega=1/P_{\rm orb}$ and the white-dwarf spin frequency by $\omega=1/P_{\rm spin}$. We regard only $\Omega$ and $\omega$ as independent physical periodicities; a peak consistent, within the frequency resolution, with an integer harmonic or a linear combination of these frequencies is classified as a harmonic or combination frequency.

We refined $\Omega$ and $\omega$ and estimated their uncertainties with a joint fit to the detrended 2-min measurements in the time domain. The model contains a Fourier representation of the non-sinusoidal orbital waveform, the spin term at $\omega$, and the beat term constrained to $\omega-\Omega$. We selected five orbital Fourier terms ($\Omega$ through $5\Omega$) by minimizing the Bayesian information criterion among orders one through ten; the terms from $3\Omega$ to $5\Omega$ describe the orbital waveform and are not treated as additional independent periodicities. The frequencies and component phases are common to both sectors, while each sector has its own constant offset and component amplitudes. This treatment preserves phase coherence across the two sectors without requiring their diluted normalized amplitudes to be identical.

After fitting the model, we performed a residual moving-block bootstrap separately within each sector, so that short-timescale correlation in the residuals and the two-sector data structure were retained. The block length was set to twice the integrated residual autocorrelation time, rounded upward, giving 30 cadences (1.00 h) for Sector 88 and 112 cadences (3.73 h) for Sector 89. For each of 5000 realizations, we added independently block-resampled sector residuals to the fitted model and refitted all model parameters. We quote the 16th and 84th percentiles of the resulting distributions as the central 68 per cent intervals. Repeating the calculation with half and twice the adopted block lengths gave consistent intervals. We also repeated the fits using 0.5-, 1.5-, and 2-d biweight windows to assess sensitivity to the adopted 1-d detrending window; the resulting period shifts are reported in Section~4.1.

For the X-ray data, we computed error-weighted generalized Lomb--Scargle periodograms, again fitting a constant offset at each trial frequency and using the standard power normalization \citep{2009A&A...496..577Z}, over periods from 60 to 7200 s. We computed a periodogram separately for each of the EPIC-pn, MOS1, and MOS2 source light curves, using 50 samples per nominal peak. The Baluev approximation \citep{2008MNRAS.385.1279B} was used to calculate the global false-alarm probability (FAP) and the nominal 1 per cent FAP threshold over the same search interval. To estimate conditional statistical period uncertainties, we generated 5000 Gaussian Monte Carlo realizations for each detector using the reported \texttt{RATE} uncertainties, preserved the original time sampling, and refitted the local frequency around each periodogram maximum. These intervals propagate the reported per-bin measurement errors conditional on the selected peak and a local sinusoidal model; they do not include correlated source variability, departures from that model, or detector-dependent systematic effects. The X-ray light curves and period-search results are presented and discussed later in Section~4.3.

\subsection{Localization of the TESS modulation}
\label{sec:localization}

We used the standard difference-image diagnostics generated by the SPOC Data Validation pipeline for the 0.22439-d Sector 88 detection. In the SPOC products, the out-of-transit image is the mean of target-pixel images outside the fitted dimming windows, whereas the in-transit image is the corresponding mean within those windows. Because the signal is associated with a CV rather than a planetary transit, we refer to these as the out-of-event and in-event images. Their difference, out of event minus in event, isolates the phase-localized flux deficit associated with the repeated dimming: sources that do not vary coherently with the event cancel approximately. SPOC fits the pixel-response function to the difference image and compares its centroid with the TIC target position. We adopted the centroid offsets reported in the SPOC Data Validation products. The centroid of the out-of-event image traces the combined mean light in the target-pixel scene and can therefore be displaced by bright neighbouring sources.

For Sector 88, the SPOC Data Validation summary reports robust-mean offsets of the fitted difference-image centroid of 50.285 arcsec ($19.86\sigma$) relative to the out-of-event image centroid and 0.558 arcsec ($0.22\sigma$) relative to the TIC target position. Because the latter result places the difference-image centroid very close to the target, the former result implies that the mean-light photocentre is displaced by approximately 50 arcsec from the target, most likely owing to flux from bright neighbouring sources. The 50.285-arcsec offset therefore does not by itself locate the periodic modulation relative to the target, whereas the 0.558-arcsec difference-centroid offset does. SPOC did not report a corresponding difference-image centroid offset for Sector 89.

As an independent check, we computed fixed-frequency Lomb--Scargle power maps for every TPF pixel at both the orbital and spin frequencies. For each pixel, we used cadences with a zero \textit{TESS} \texttt{QUALITY} bit mask and valid time and pixel-flux measurements. Slow variations were removed with a 1-d biweight filter applied independently to each pixel within each sector. We then evaluated the Lomb--Scargle power at the measured frequencies, $\Omega=4.45650$ d$^{-1}$ and $\omega=34.20962$ d$^{-1}$. The Gaia sky coordinates were transformed to floating-point TPF coordinates using the FITS world-coordinate system (WCS). One TPF grid square is one physical \textit{TESS} pixel (approximately 21 arcsec), while the pixel-response function distributes a point source over several pixels; the WCS position therefore gives the continuous position of the source centroid within the pixel grid.

\section{Results and Discussion}
\label{sec:results}

\subsection{TESS frequencies}

\begin{figure*}[htbp]
\centering
\includegraphics[width=1.0\textwidth]{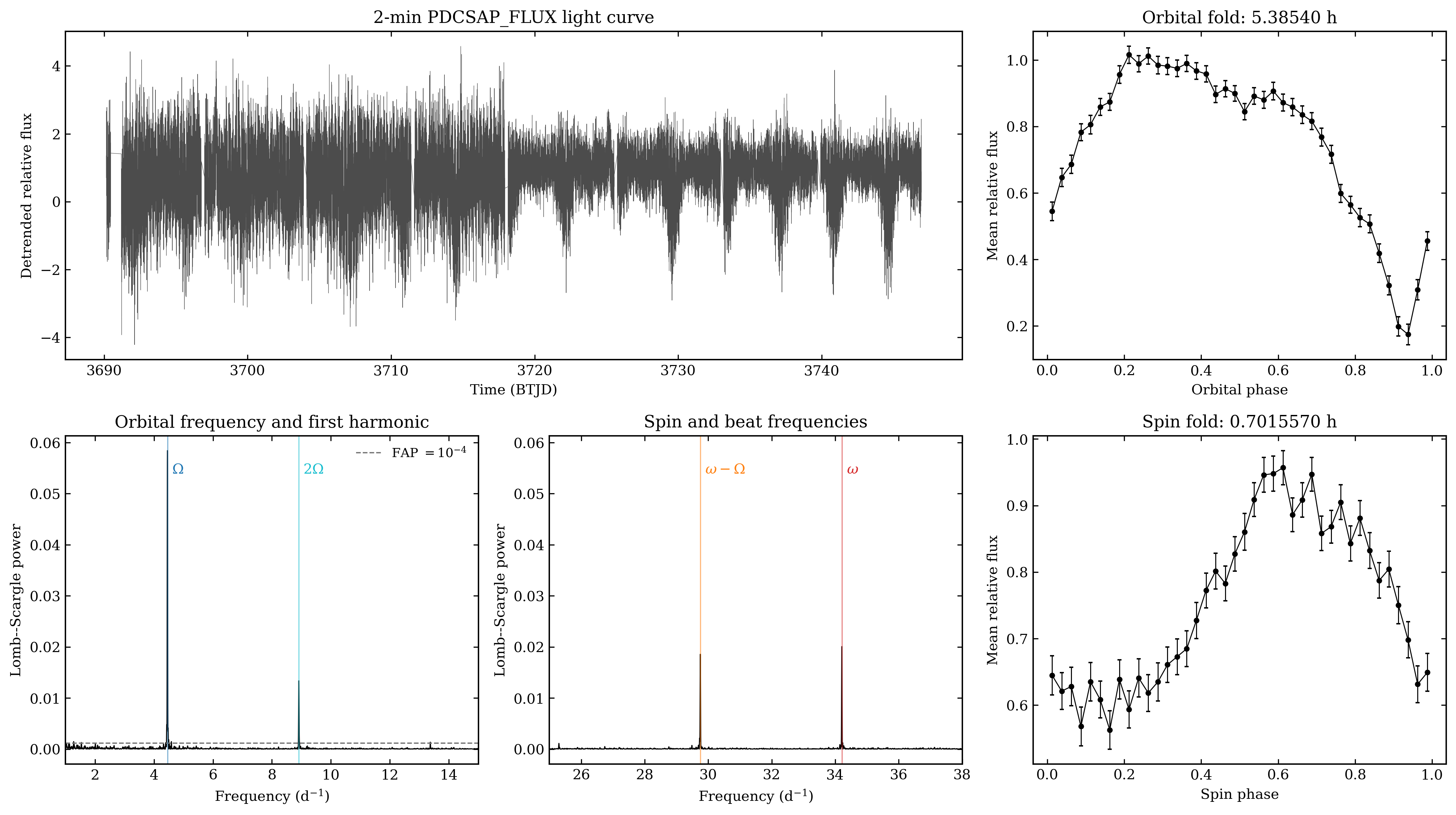}
\caption{\textit{TESS} timing analysis of Gaia 5533 using the concatenated, sector-normalized 2-min \texttt{PDCSAP\_FLUX} data. Top: the detrended light curve (left) and the 5.38540-h orbital phase profile (right). Bottom: the orbital frequency $\Omega$ and first harmonic $2\Omega$ (left), the white-dwarf spin frequency $\omega$ and spin--orbit beat frequency $\omega-\Omega$ (middle), and the 0.7015570-h spin phase profile (right). Both phase profiles use 40 equal-width bins for display; the periodograms and time-domain fit use the original 2-min measurements.}
\label{fig:tess_timing}
\end{figure*}

Figure~\ref{fig:tess_timing} shows four prominent, physically related signals. The joint time-domain fit gives $\Omega=4.45650^{+0.00023}_{-0.00022}$ d$^{-1}$, or $P_{\rm orb}=5.38540^{+0.00027}_{-0.00028}$ h. The peak at $2\Omega=8.91299$ d$^{-1}$ is the first orbital harmonic and corresponds to 2.69270 h. The high-frequency signal gives $\omega=34.20962^{+0.00030}_{-0.00029}$ d$^{-1}$, or $P_{\rm spin}=0.7015570^{+0.0000059}_{-0.0000062}$ h. The remaining prominent peak agrees with the constrained combination frequency $\omega-\Omega=29.75313^{+0.00031}_{-0.00030}$ d$^{-1}$, corresponding to an optical beat period of $0.8066379^{+0.0000082}_{-0.0000084}$ h. The quoted uncertainties are central 68 per cent bootstrap intervals. The $2\Omega$ and $\omega-\Omega$ signals are interpreted as the first orbital harmonic and the spin--orbit beat frequency, respectively, rather than as additional independent periods.

The 56.792-d two-sector baseline corresponds to a Rayleigh frequency resolution of $1/T=0.01761$ d$^{-1}$. The period uncertainties quoted above were determined from the bootstrap distributions. Repeating the analysis with half and twice the adopted residual block lengths produced consistent intervals, while varying the biweight detrending window from 0.5 to 2 d shifted $P_{\rm orb}$ by at most 0.11 s and $P_{\rm spin}$ by at most 0.0013 s, both smaller than the bootstrap uncertainties.

The 5.38540-h modulation is detected separately in Sectors 88 and 89 and is also present in both the 20-s and 2-min cadence products. The combined 2-min \texttt{PDCSAP\_FLUX} result is consistent with the 0.22439-d period reported in the Sector 88 SPOC Data Validation product. When folded on 5.38540 h, measurements from successive cycles align to form one dominant, broad, asymmetric depression per cycle, with a relatively sharp minimum and no comparably strong secondary minimum. The coherent recurrence of this feature supports the 5.38540-h periodicity. Its non-sinusoidal morphology naturally produces Fourier power at integer harmonics. We therefore identify the $2\Omega$ peak as the first harmonic of the 5.38540-h modulation rather than as an independent periodicity. Together with the observed $\omega-\Omega$ frequency relation, these properties support the identification of $\Omega$ as the orbital frequency.

\subsection{Spatial origin of the orbital and spin modulations}

\begin{figure*}[htbp]
\centering
\includegraphics[width=1.0\textwidth]{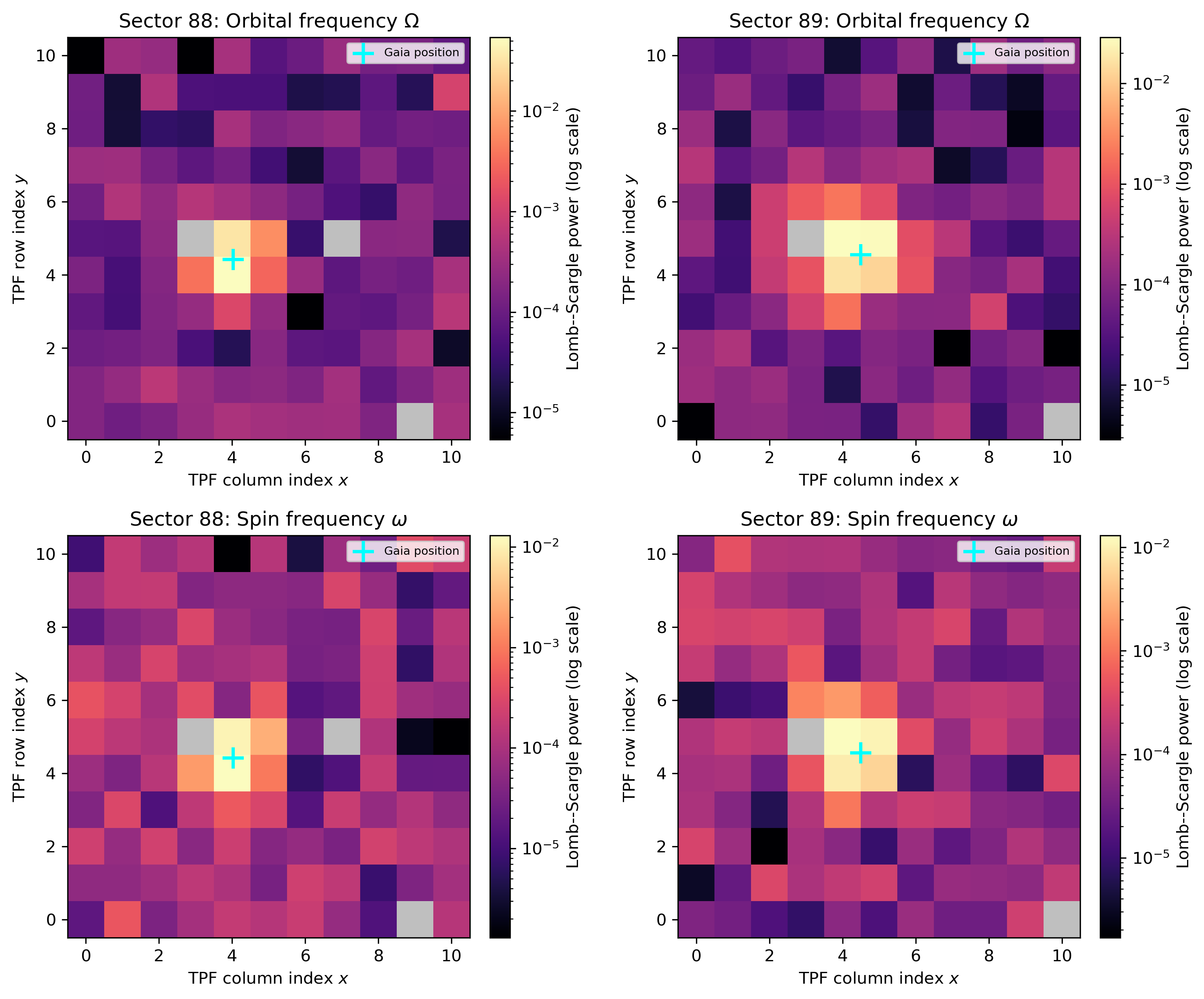}
\caption{Fixed-frequency Lomb--Scargle power maps for the TPF pixels in Sectors 88 and 89. The top row gives the power at the orbital frequency $\Omega$ and the bottom row the power at the spin frequency $\omega$. Colour represents the Lomb--Scargle power at the indicated frequency, and the cyan plus marks the Gaia position transformed through the TPF WCS. Each coloured square is one physical \textit{TESS} pixel.}
\label{fig:pixel_power}
\end{figure*}

Figure~\ref{fig:pixel_power} shows the fixed-frequency power maps at the orbital and spin frequencies. The WCS target positions are $(x,y)=(4.034,4.421)$ in Sector 88 and $(4.497,4.556)$ in Sector 89. At both frequencies, the strongest power is concentrated in the pixel-response-function footprint surrounding these positions in both sectors. These maps provide a pixel-level localization check that complements the SPOC difference-image centroid.

Although the low \texttt{CROWDSAP} values indicate severe dilution of the mean aperture flux, they do not determine the spatial origin of the periodic signal. In Sector 88, the out-of-event photocentre is offset from the target by 50.285 arcsec because the mean target-pixel scene is dominated by neighbouring flux. In contrast, the difference-image centroid of the 5.38540-h event lies only 0.558 arcsec from Gaia 5533, corresponding to 0.22 standard deviations. The closest catalogued neighbour in the SPOC report, TIC 821425373, is 8.79 arcsec from the target. The orbital- and spin-frequency pixel-power maps in both sectors further localize the signals to the pixel-response-function footprint around the WCS target position. These diagnostics identify Gaia DR3 5533020382479508736 as the spatial source of the measured TESS modulations.

\subsection{X-ray spin modulation and IP classification}

\begin{figure*}[htbp]
\centering
\includegraphics[width=1.0\textwidth]{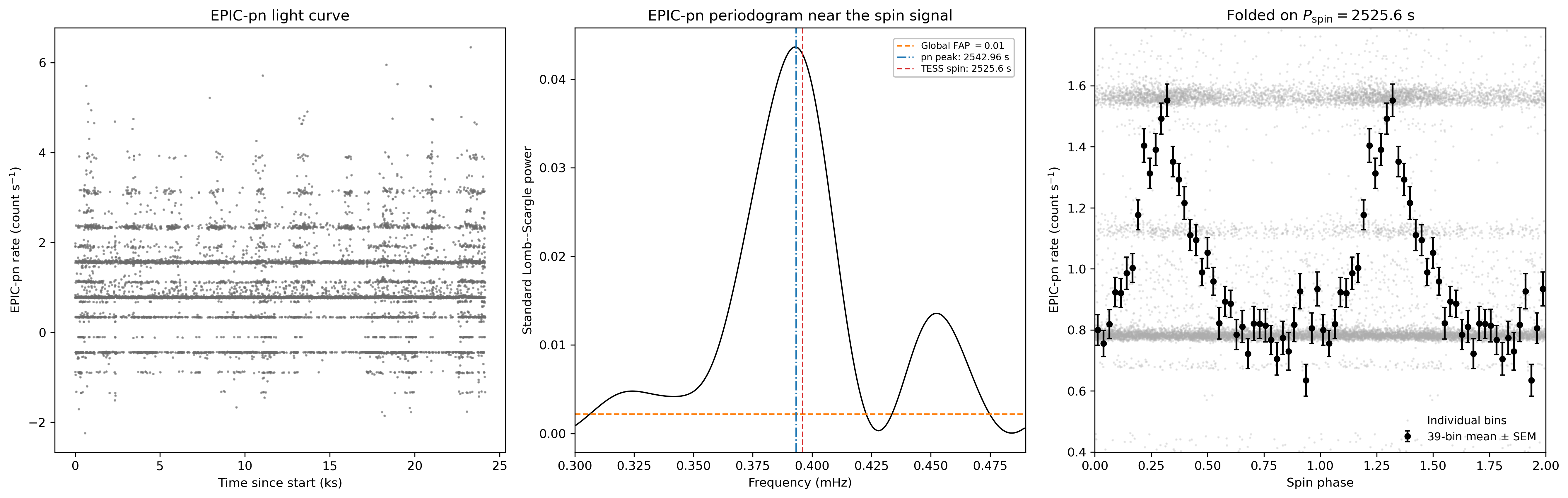}
\caption{Archival \textit{XMM-Newton} EPIC-pn timing analysis of Gaia 5533. Left: the source light curve. Middle: the standard-normalized Lomb--Scargle periodogram zoomed to the spin-signal region; the orange dashed line marks the power threshold corresponding to a nominal global FAP$=0.01$ calculated with the Baluev approximation over 60--7200 s, the blue dash-dotted line marks the EPIC-pn maximum at 2542.96 s, and the red dashed line marks the $2525.6$-s \textit{TESS} spin period. Right: the EPIC-pn light curve folded on the optical spin period and repeated over two cycles. Black points show the mean count rate in 39 equal-width phase bins with standard errors of the means; gray points show individual measurements. Phase binning is used only for display.}
\label{fig:xray}
\end{figure*}

The periodogram maxima occur at 2542.96, 2557.57, and 2487.91 s in the EPIC-pn, MOS1, and MOS2 source light curves, respectively. Their standard-normalized peak powers are 0.04363, 0.30131, and 0.37442, and the corresponding global Baluev FAPs are $1.52\times10^{-101}$, $1.79\times10^{-64}$, and $2.99\times10^{-83}$. The nominal global FAP$=0.01$ levels are 0.00220, 0.02703, and 0.02725, respectively, so each maximum is highly significant under the adopted analytic test. Local sinusoidal fits give 2545.17, 2556.02, and 2486.76 s. The conditional 68 per cent intervals from 5000 Gaussian Monte Carlo realizations are $2544.89^{+6.84}_{-6.45}$ s, $2555.92^{+3.22}_{-3.15}$ s, and $2486.81^{+3.02}_{-3.14}$ s for EPIC-pn, MOS1, and MOS2, respectively. These conditional intervals do not overlap, showing that the detector-to-detector peak shifts exceed the per-bin statistical errors propagated by the simulations. Because the simulations do not include correlated source variability, departures from the local sinusoidal model, or detector-dependent systematic effects, we do not interpret the non-overlap as evidence for three physical periods. The three periodogram maxima lie within 39 s (1.5 per cent) of the 2525.6-s ($0.701557$ h) optical spin period. Their separations are also smaller than the local Rayleigh period scales of approximately 244--264 s set by the usable wall-clock baselines near 2526 s; these scales describe the broad peak widths expected from the short observing baselines. We therefore interpret the three significant X-ray features conservatively as broad peaks near the optical spin period, without claiming that their detector-specific best-fitting periods are identical. Figure~\ref{fig:xray} shows, from left to right, the EPIC-pn X-ray light curve, its standard-normalized Lomb--Scargle periodogram, and the same light curve phase-folded at the optical spin period. As the magnetically channelled accretion regions rotate with the WD, their projected visibility and the line-of-sight absorption through the accretion column can vary with spin phase, producing an X-ray modulation at the spin frequency. The presence of significant X-ray power in the same period region supports the identification of the 0.7015570-h optical signal with the white-dwarf spin period. The $\omega-\Omega$ peak is the spin--orbit beat frequency. Previous work established the \textit{Swift}/BAT hard-X-ray detection and the optical CV classification, and favoured the XRT source coincident with the CV as the X-ray counterpart \citep{2010A&A...524A..64C,Rojas2017,Landi2017}. Although the optical line ratio had led \citet{Rojas2017} to suggest a possible polar classification, the measured ratio $P_{\rm spin}/P_{\rm orb}=0.1303$ demonstrates strong asynchronism and distinguishes the system from a synchronized polar. Together with the previously published properties, the new \textit{TESS} and \textit{XMM-Newton} timing results identify 2PBC J0800.5$-$4306 as an IP.

\section{Conclusion}
\label{sec:conclusion}

We have determined the previously unknown orbital and white-dwarf spin periods of 2PBC J0800.5$-$4306, a hard-X-ray-selected CV whose magnetic subtype had remained uncertain. The combined Sector 88 and Sector 89 \textit{TESS} analysis yields $P_{\rm orb}=5.38540^{+0.00027}_{-0.00028}$ h and $P_{\rm spin}=0.7015570^{+0.0000059}_{-0.0000062}$ h; the additional peaks at $2\Omega$ and $\omega-\Omega$ are identified as the first orbital harmonic and the spin--orbit beat frequency, respectively. The archival \textit{XMM-Newton} EPIC-pn, MOS1, and MOS2 light curves show broad periodogram peaks at 2543, 2558, and 2488 s, respectively, near the optical spin period. Together with the previously published CV classification and \textit{Swift}/BAT hard-X-ray detection, these timing results identify 2PBC J0800.5$-$4306 as an intermediate polar. SPOC difference-image diagnostics and pixel-level timing tests further support that the measured \textit{TESS} modulations are associated with the Gaia counterpart despite the crowded field.

\begin{acknowledgements}
This work received the support of the National Natural Science Foundation of China under grants 12463011, 12563007, 12373038, 12541303 and 12288102; the Natural Science Foundation of Xinjiang No. 2024D01C230 and 2022D01D85; and the China Manned Space Program under grant No. CMS-CSST--2025-A15.

This paper includes data collected by the TESS mission. Funding for the TESS mission is provided by NASA's Science Mission Directorate. This research also makes use of data obtained from \textit{XMM-Newton}, an ESA science mission with instruments and contributions directly funded by ESA Member States and NASA, and of the SIMBAD database \citep{2000A&AS..143....9W}, operated at CDS, Strasbourg, France.

Software: Astropy \citep{2013A&A...558A..33A,2018AJ....156..123A,2022ApJ...935..167A}, lightkurve \citep{2018ascl.soft12013L}, wotan \citep{2019AJ....158..143H}, Matplotlib \citep{2007CSE.....9...90H}, NumPy \citep{2020Natur.585..357H}, SciPy \citep{2020NaMet..17..261V}.
\end{acknowledgements}

\bibliographystyle{raa}
\bibliography{ms2026-0594}

\label{lastpage}
\end{document}